\documentclass[letter,longauth]{aa} 
\usepackage{graphicx}
\usepackage{txfonts}
\usepackage{natbib}
\bibpunct{(}{)}{;}{a}{}{,} 
\usepackage{color}
\newcommand{\um}{$\mu$m}                                 
\newcommand{\msun}{$M_{\odot}$}

\newcommand{\gapprox}{$\stackrel {>}{_{\sim}}$}   
\newcommand{\lapprox}{$\stackrel {<}{_{\sim}}$}
\newcommand{\about}{$\sim$}                       
\newcommand{\qeri}{q$^{1}$\,Eri}

\newcommand{\asec}{$^{\prime \prime}$}
\newcommand{\adeg}{$^{\circ}$}

\newcommand{\radot}[4]{\mbox{#1$^{\rm h}$#2$^{\rm m}$#3$\stackrel{\rm s}
{_{\bf\cdot}}$#4}}  

\newcommand{\decdot}[4]{\mbox{#1$^{\circ}$ #2$^{\prime}$ #3$\stackrel {\prime 
\prime}{_{\bf \cdot}}$#4}}

\newcommand{\asecdot}[2]{\mbox{#1$\stackrel {\prime \prime}{_{\bf \cdot}}$#2}}

\begin{document}
%
   \title{Resolving the cold debris disc around a planet-hosting star\thanks{{\it Herschel} is an ESA space observatory with science instruments provided by European-led Principal Investigator consortia and with important participation from NASA.}}
   
   \subtitle{PACS photometric imaging observations of q$^{1}$\,Eri (HD\,10647, HR\,506)}

   \author{R. Liseau\inst{1}
	\and
	         C. Eiroa\inst{2}
	 \and	
          	D. Fedele\inst{2,\,3,\,4}
	\and
		J.-C. Augereau\inst{5}
	\and
		G. Olofsson\inst{6}	
         	 \and
		B. Gonz\'{a}lez\inst{7}
	\and
 		J. Maldonado\inst{2}
	\and
		B. Montesinos\inst{8}
	\and
		A. Mora\inst{2,\,9}
	\and
		O. Absil\inst{10}
	\and 
		D. Ardila\inst{11} 
	\and
		 D. Barrado\inst{8,\,12}
	\and	
		A. Bayo\inst{8,\,24}
	\and
		C.A. Beichman\inst{13}
	\and	
		G. Bryden\inst{14}
	\and
		W. C. Danchi\inst{15}
	\and
		C. del Burgo\inst{16}		
	\and 
		S. Ertel\inst{17}
	 \and
	  	C.W.M. Fridlund\inst{18}
	 \and
	  	A.M. Heras\inst{18}
	\and 	
		A.V. Krivov\inst{19}	
	\and
		R. Launhardt\inst{3}
	\and
		J. Lebreton\inst{5}
	\and
		T. L\"{o}hne\inst{19}
	\and	
		J.P. Marshall\inst{20}
	 \and
	 	G. Meeus\inst{2}
	 \and
	 	S. M{\" u}ller\inst{19}
	 \and
	          G.L. Pilbratt\inst{18} 
	 \and
	 	A. Roberge\inst{15}
	\and	
	  	J. Rodmann\inst{21}  
	 \and
	 	E. Solano\inst{8}
	 \and
		K.R. Stapelfeldt\inst{14}
	 \and
		Ph.Th\'{e}bault\inst{22}
         	 \and
		G.J. White\inst{20,\,23}
	\and
		S. Wolf\inst{17}	
	  }
    
   \institute{Onsala Space Observatory, Chalmers University of Technology, S-439 92 Onsala, Sweden, \email{rene.liseau@chalmers.se}
          \and
             	Departamento de F\'{i}sica Te\'{o}rica, C-XI, Facultad de Ciencias, Universidad Aut\'{o}noma de Madrid, Cantoblanco, 28049 Madrid, Spain
	\and 
		Max Planck Institut f\"{u}r Astronomie, K\"{o}nigstuhl 17, 69117 Heidel-\\
		berg, Germany
         \and
		Johns Hopkins University Dept. of Physics and Astronomy, 3701 San Martin drive Baltimore, MD 21210 USA 
	\and
        		Universit\'{e} Joseph Fourier/CNRS, Laboratoire d'Astrophysique de Grenoble, UMR 5571, Grenoble, France   
         \and
             	Stockholm Observatory, AlbaNova University Center, Roslagstulls-\\
		backen 21, SE-106 91 Stockholm, Sweden     
	\and
		INSA at ESAC, E-28691 Villanueva de la Ca\~{n}ada, Madrid, Spain
	\and	
		LAEX-CAB, Depto. Astrof\'{\i}sica, Centro de Astrobiolog\'{\i}a (INTA-CSIC), P.O. Box 78,   E-28691 Villanueva de la Ca\~nada, Spain
        	\and
		ESA-ESAC Gaia SOC. P.O. Box 78, E-28691 Villanueva de la Ca{\~n}ada, Madrid, Spain
         \and
		Institut d'Astrophysique et de G{\'e}ophysique, Universit{\'e} de Li{\`e}ge, 17 All{\'e}e du Six Ao{\^u}t, B-4000 Sart Tilman, Belgium
	\and
		{\it Spitzer} Science Center, California Institute of Technology, Pasadena, CA 91125, USA
	\and
		Centro Astron\'omico Hispano Alem\'an de Calar Alto (CAHA), C/ Jes\'us Durb\'an Remo\'n 2-2 E-04004 Almer\'{\i}a, Spain	
	\and
		NASA Exoplanet Science Institute, California Institute of Technology, Pasadena, CA 91125, USA
	\and
		Jet Propulsion Lab, California Institute of Technology, 4800 Oak Grove Dr, Pasadena, CA 91109, USA
	\and
		NASA Goddard Space Flight Center, Exoplanets and Stellar Astrophysics, Code 667, Greenbelt, MD 20771
	\and
		UNINOVA-CA3, Campus da Caparica, Quinta da Torre, Monte de Caparica, 2825-149 Caparica, Portugal
	\and
		University of Kiel, Institute of Theoretical Physics and Astrophysics, Leibnizstrasse 15, 24098 Kiel, Germany
	\and		
	    	ESA Astrophysics \& Fund Physics Missions Division, ESTEC/SRE-SA, Keplerlaan 1, NL-2201 AZ Noordwijk, The Netherlands
	\and
		Astrophysikalisches Institut und Universit{\"a}tssternwarte, Friedrich-Schiller-Universit{\"a}t, Schillerg{\"a}{\ss}chen 2-3, 07745 Jena, Germany
	\and 
		Department of Physics and Astronomy, Open University, Walton Hall, Milton Keynes MK7 6AA, UK 
	\and
		ESA/ESTEC Space Environment and Effects Section, PO Box 299, 2200 AG Noordwijk, The Netherlands
         \and	 
		LESIA, Observatoire de Paris, 92195, Meudon, France
	\and
		Science and Technology Facilities Council, Rutherford Appleton Laboratory, Chilton, Didcot OX11 0QX, UK 
	\and
		European Southern Observatory, Alonso de C\'{o}rdova 3107, Vitacura, Santiago, Chile	
	}

   \date{Received ; accepted }


  \abstract
   {About two dozen exo-solar debris systems have been spatially resolved.  These debris discs commonly display a variety of structural features such as clumps, rings, belts, excentric distributions and spiral patterns. In most cases, these features are believed to be formed, shaped and maintained by the dynamical influence of planets orbiting the host stars. In very few cases has the presence of the dynamically important planet(s) been inferred from direct observation.}
   {The solar-type star \qeri\ is known to be surrounded by debris, extended on scales of \lapprox\,30\asec. The star is also known to host at least one planet, albeit on an orbit far too small to make it responsible for structures at distances of tens to hundreds of AU. The aim of the present investigation is twofold: to determine the optical and material properties of the debris and to infer the spatial distribution of the dust, which may hint at the presence of additional planets.}
   {The photodetector array camera and spectrometer (PACS) aboard the {{\it Herschel}} Space Observatory allows imaging observations in the far infrared at unprecedented resolution, i.e. at better than 6\asec\ to 12\asec\ over the wavelength range of 60\,\um\ to 210\,\um. Together with the results from ground-based observations, these spatially resolved data can be modelled to determine the nature of the debris and its evolution more reliably than what would be possible from unresolved data alone.}
   {For the first time has the \qeri\ disc been resolved at far infrared wavelengths. The PACS observations at 70\,\um, 100\,\um\ and 160\,\um\ reveal an oval image showing a disc-like structure in all bands, the size of which increases with wavelength. Assuming a circular shape yields the inclination of its equatorial plane with respect to that of the sky, $i > 53$\adeg. The results of image de-convolution indicate that $i$ likely is larger than 63\adeg, where 90\adeg\ corresponds to an edge-on disc.}
   {The observed emission is thermal and optically thin. The resolved data are consistent with debris at temperatures below 30\,K at radii larger than 120\,AU. From image de-convolution, we find that \qeri\ is surrounded by an about 40\,AU wide ring at the radial distance of \about\,85\,AU. This is the first real Edgeworth-Kuiper Belt analogue ever observed.}

   \keywords{Stars: individual: q$^{1}$\,Eri (HD\,10647, HR\,506, HIP\,7978) -- 
             Stars: planetary  systems: planetary discs -- circumstellar matter -- formation
               }

   \maketitle
%

\section{Introduction}

We report new observations of the debris disc around \qeri\ (HD\,10647, HR\,506) using the {\it Herschel} Space Observatory \citep{pilbratt2010}. The observations form part of a larger Key Programme (KP), viz. DUNES\footnote{DUst around NEarby Stars, {\tiny {\ttfamily http://www.mpia-hd.mpg.de/DUNES/}}}, which is described in more detail by \citet{eiroa2010}. Here, we give a brief summary to put the contents of this Letter into context. The DUNES KP is a sensitivity limited study with the goal of discovering and characterising extra-solar analogues of the Edgeworth-Kuiper Belt (EKB) in an unbiased, statistical sample of nearby F, G and K main-sequence stars. The sample is volume limited, with distances \lapprox\,20\,pc, and spans a broad range of stellar ages, from \about\,0.1 to roughly 10\,Gyr. In addition to the object of the present study (\qeri), a number of M- and A-type stars will be observed in collaboration with the DEBRIS-KP team \citep{matthews2010}, implying that the whole sample covers a decade in stellar mass from 0.2 to 2\,\msun.

The PACS \citep{poglitsch2010} observations at 100\,\um\ aim at the detection of the stellar photospheres down to the confusion noise with a signal to noise ratio (S/N) of at least 5. Together with observations in the other {\it Herschel} bands, this will lead to an unprecedented characterisation of discs and will allow detailed theoretical modelling. It is foreseen that fractional luminosities $L_{\rm dust} / L_{\odot}$ of a few times $10^{-7}$ will be reached, i.e. similar to that of the EKB of the Solar System \citep{stern1996,jewitt2009} and a more than order of magnitude improvement over {\it Spitzer} data \citep{bryden2009}.

\begin{figure*}[t]
  \resizebox{\hsize}{!}{
  \rotatebox{00}{\includegraphics{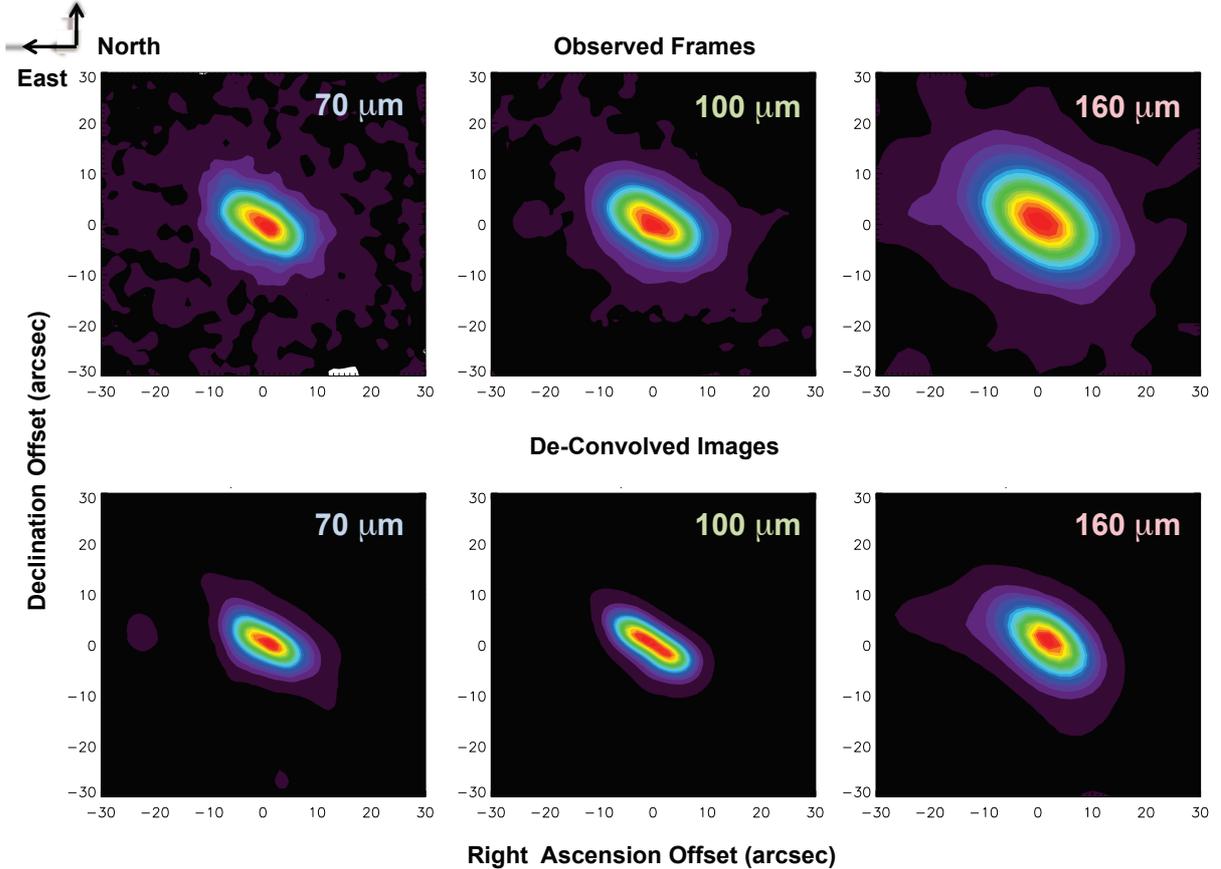}}
  }
  \caption{PACS photometric imaging of \qeri\ at, from left to right, 70\,\um\  ({\it blue}), 100\,\um\ ({\it green}) and 160\,\um\ ({\it red}). The 70\,\um\ image was taken in chop-nod mode, whereas the other two in scan-map mode. The upper panels display the reduced observations. Below, de-convolved images are shown, using observations of $\alpha$\,Boo for the definition of the PSF. Displayed are the results for ten iterations of a MEM algorithm \citep{hollis1992}. The star defines the origin of the frames, i.e. offset coordinates (0, 0). Within the positional accuracy (2\asec\ rms), the stellar position and the centre of the elliptical brightness distributions coincide (see Table\,\ref{results}) and offsets are in seconds of arc. The lowest contours are at 5\% of the maximum values and consecutive steps are also by this amount. At the distance of the star, 20\asec\ corresponds to 350\,AU.
  }
  \label{observed}
\end{figure*}

A main-sequence star with an infrared excess larger by more than three orders of magnitude than this limit is the late F-type star \qeri. This will potentially allow to study the material giving rise to the excess in great detail. The star is known to be accompanied by a giant planet, \qeri\,b, orbiting at the distance of 2\,AU \citep[][and references therein] {butler2006}, which corresponds to \asecdot{0}{1} at the distance of the system \citep[17.35\,pc;][]{perryman1997,vanleeuwen2007}. Such a small angle will not be resolved by the observations described in this paper, but \qeri\ is also surrounded by a ring or belt system of scattering and thermally emitting dust particles. The extent of the system, which from optical to sub-millimetre wavelengths is up to several tens of arc seconds in size \citep[e.g.,][and references therein]{stapelfeldt2007,liseau2008}, should be readily accessible to PACS at the {\it Herschel} telescope ($\theta_{\rm diff}^{\,\prime \prime} = 7 \times \lambda_{\mu{\rm m}}/100\,\mu{\rm m}$).

This contribution presents the observed properties of the \qeri\ system by {\it Herschel}. The results of theoretical model calculations will be communicated by Augereau et al. (in prep.). The observations and data reduction are described in Sect.\,2, with the results presented in Sect.\,3. These are discussed in Sect.\,4 and, finally in Sect.\,5, our main conclusions are briefly summarised.

\begin{table*}
\begin{flushleft}
 \caption{\label{results} Brightness, orientation and extent of the \qeri\ debris system}
\resizebox{\hsize}{!}{
\begin{tabular}{lcllllll}
  \hline
  \noalign{\smallskip}
Observation & $\Delta {\rm RA}$, $\Delta {\rm Dec}^{a}$ 	& Flux$^{b}$			& 2a$^{b}$	&   2b$^{b}$			& PA$^{c}$     	& $i^{\,d}$		& Alternate flux		\\
 ID and Mode    		&	(\asec, \asec)			& (mJy)				&(\asec)		&  (\asec)				& (\adeg) 		& (\adeg)		& measurement (mJy)	\\
\noalign{\smallskip}
\hline \\                                 
1342187142 ~~ - \phantom{1}70 :  chop-nod&$+0.9$, $-0.9$& $ 828 \pm 83$ 		& 13.9 (11.0)	& \phantom{1}7.9 (4.4)	&   54.4 (60.0)	& 55.4 (66.4)	& \phantom{1}$859 \pm  \phantom{1}6$ {\it Spitzer}/MIPS$^{e}$\\
1342187141~~~ - 100 :  chop-nod	 &   $+2$, $0$		& $ 816 \pm 82$		& 15.5 (12.6)	& \phantom{1}8.8 (5.1) 	&   54.4 (56.2)	& 55.4 (66.1)	& $1080\pm 36$ IRAS FSC\\
134218739/40 - 100 : scan map	&     $0$, $-1.4$	& $ 810 \pm 81$		& 15.1 (12.5)	& \phantom{1}8.8 (4.7)	&   56.1 (54.8)	& 54.4 (67.9)	& $1080\pm 36$ IRAS FSC\\
1342187142 ~~~- 160 :  chop-nod 	& $-1.4$, $+0.9$	& $ 529 \pm 106$		& 18.9 (13.5)	& 12.3 (7.8)			&   51.3 (48.9)	& 49.4 (54.7)	& \phantom{1}$453 \pm 50$ {\it Spitzer}/MIPS$^{f}$\\
134218739/40 - 160 : scan map	&$+1.4$, $-1.6$	& $ 537 \pm 107$		& 19.3 (13.3)	& 12.5 (6.5)	 		&   51.2 (55.7)	& 49.6 (60.7)	& \phantom{1}$453 \pm 50$ {\it Spitzer}/MIPS$^{f}$\\	
  \noalign{\smallskip}
   \hline
  \end{tabular}
   }
\end{flushleft}
Notes to the Table: \\
$^{a}$ Offsets of ellipse centre relative to stellar coordinates: proper motion corrected to IRCS 2009.89 \citep[+0.186\,s and \asecdot{$-1$}{055}][]{perryman1997,vanleeuwen2007} \radot{01}{42}{29}{502}, \decdot{$-53$}{44}{28}{06}. The photospheric fluxes are 17\,mJy  at 70\,\um, 8\,mJy at 100\,\um\ and 3\,mJy at 160\,\um\ (cf. Fig.\,\ref{SED}). \\
$^{b}$ Flux for Gaussian ellipse with fitted FWHM to major and minor axis, 2a and 2b, respectively. Values in parentheses refer to de-convolved images.\\ 
$^{c}$ Position angle measured from North over East.\\
$^{d}$ Lower limit to the inclination, where $i=90$\adeg\ refers to an edge-on geometry.\\ 
$^{e}$ \citet{trilling2008}.\\ 
$^{ f}$ \citet{tanner2009}. 
\end{table*}

\section{Observations and data reduction}

During this initial observing run, two different modes of observing were executed for test reasons, in order to optimise the efficiency of the programme in terms of observing time and signal-to-noise ratio (S/N). These modes were, respectively, the chop-and-nod mode (70, 100 and 160\,\um), adopted for point source observing, and the scan-map option (100 and 160\,\um), for extended sources (Fig.\,\ref{observed}). These modes are described by \citet{eiroa2010} and in a future technical note.

The reduction was done within the {\it Herschel} interactive processing environment, using HIPE\_v2.0.0\_RC3, and with scripts for the pipeline developed by members of the PACS-ICC\footnote{{\it Herschel} science demonstration phase data processing workshop, 14-16 December 2009, ESAC, Spain.}.  At the medium scan speed of 20\asec\,s$^{-1}$, two maps along position angles 63\adeg\ and 117\adeg, respectively, were obtained in order to minimise striping in the resultant images. At both 100\,\um\ and 160\,\um, the sky noise was lower in the scan-map data (2.8 and $5.2) \times 10^{-5}$\,Jy\,arcsec$^{-2}$, respectively, as compared to (3.9 and $12) \times 10^{-5}$\,Jy\,arcsec$^{-2}$ of the chop-nod observations. These values are comparable to the surface brightness of the scattering disc at 0.6\,\um\ \citep[$1.3 \times 10^{-5}$\,Jy\,arcsec$^{-2}$, as deduced from data in ][]{stapelfeldt2007}.

The prime calibrator Arcturus \citep[$\alpha$ Boo; e.g.,][]{cohen2005} was observed close in time with \qeri\ and was used to provide the instrumental point spread function (PSF)\footnote{See
{\tiny {\ttfamily {http://Herschel.esac.esa.int/AOTsReleaseStatus.shtml}}} \\
{\tiny {\ttfamily {http://Herschel.esac.esa.int/Docs/AOTsReleaseStatus/PACS\_Scan\\
ChopNod\_ReleaseNote\_22Feb2010.pdf}}}  \\
{\tiny {\ttfamily {http://Herschel.esac.esa.int/Docs/AOTsReleaseStatus/PACS\_Phot\\
Map\_ReleaseNote\_23Feb2010.pdf}}}  }.

The filters at the reference wavelengths of 70, 100 and 160\,\um\ are referred to as {\it blue}, {\it green} and {\it red}, spanning $60-85$\,\um, $85-130$\,\um\ and $130-210$\,\um, respectively \citep{poglitsch2008}. The PSF has a tripolar shape at low intensities, i.e. a few percent of the peak value, and the half-power-width is circular for the 70\,\um\ and 100\,\um\  filters, but somewhat elongated at 160\,\um\ in the scan direction. Currently, the estimated accuracy of the absolute flux calibration is estimated to be better than 20\% in the long-wave and better than 10\% in the short-wave bands.

\section{Results}

In the Scan Map AOT release is the following note: {\it The fluxes are too high and have to be scaled down by the following factors 1.05 in the blue band, 1.09 in the green band 1.29 in the red band}, and in Table\,\ref{results}, no colour correction has been applied to the reported fluxes. The internal consistency of the PACS data - at a given wavelength but for different observing modes - is strikingly good and lends confidence to the quality of these data. From the comparison with previous measurements, it is apparent that IRAS fluxes are on the high side, whereas long-wave {\it Spitzer} data are somewhat on the low side. 

At the level of the pointing accuracy of {\it Herschel} (2\asec\ rms, corresponding to 35\,AU), there is no significant offset of the centres of the elliptical isophotes with respect to the position of the star (Table\,\ref{results}). This uncertainty is much larger than the offset of 8\,AU observed for Fomalhaut \citep{kalas2008}. For the nominal wavelengths of 70\,\um, 100\,\um\ and 160\,\um, the angular resolution is limited to about 6\asec, 7\asec\ and \asecdot{11}{5}, respectively. The image of \qeri\ seems clearly resolved, therefore, only in the long dimension. 

The average of the position angle of the elliptical source is PA = 54\adeg\ and the average of the lower limit to the inclination of an intrinsically circular feature (ring or belt) is $i > 53$\adeg, with estimated  uncertainties of $\sim \pm 5$\adeg\ ($i=90$\adeg\ for an edge-on geometry).  For the de-convolved images (see below), the disc would be seen more edge-on and in this case, $i > 63$\adeg. This is consistent with the tilt derived from optical images of the light scattered off the disc \citep[$i=76$\adeg,][]{stapelfeldt2007}.

\section{Discussion}

\begin{figure}[t]
  \resizebox{\hsize}{!}{
  \rotatebox{00}{\includegraphics{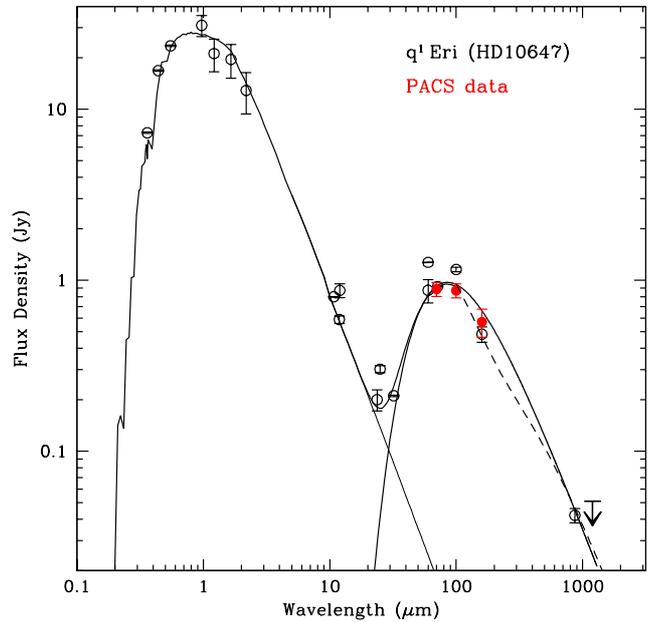}}
  }
  \caption{This figure is similar to the SED of \qeri\ shown in \citet{liseau2008} with the difference that the PACS data of Table\,\ref{results} are shown by the filled symbols. In addition, the 160\,\um\ flux of {\it Spitzer} \citep{tanner2009} has also been added. As before, the solid curve for the IR excess is that of a single temperature blackbody ($\beta = 0$, where $\kappa_{\nu} \propto \nu^{\,\beta}$), and the dashed line refers to the ring-belt composite model of that paper.
  }
  \label{SED}
\end{figure}

\begin{figure}[t]
  \resizebox{\hsize}{!}{
  \rotatebox{00}{\includegraphics{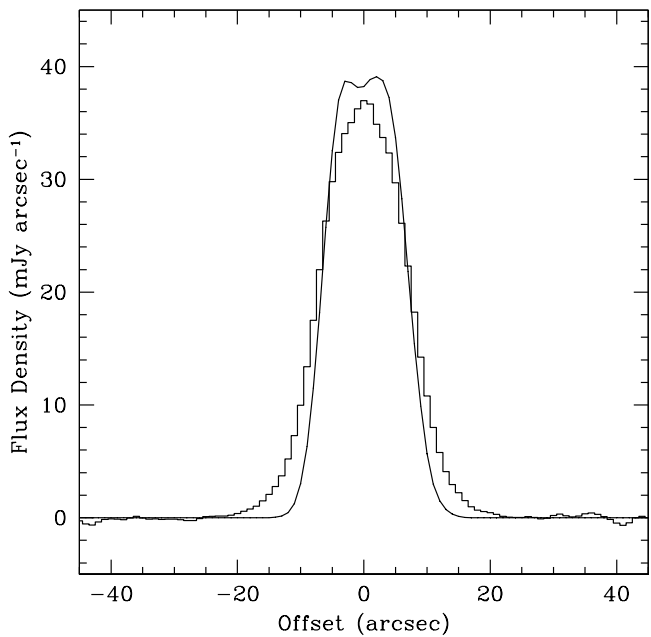}}
  }
  \caption{One-dimensional cuts (averages of 5 pxl wide strips), along the major axis, through the 100\,\um\ scan map image, from which the stellar source has been subtracted prior to the de-convolution. The histogram depicts the observed data, whereas the smooth line shows the result of the applied de-convolution algorithm.  
  }
  \label{ring}
\end{figure}

\subsection{The spectral energy distribution (SED) revisited}

Figure\,\ref{SED} is taken from the paper by \citet{liseau2008}, but with the PACS data included. The 70\,\um\ and 100\,\um\ fluxes fall essentially on top of the 60\,K blackbody curve, shown as a solid line. For 160\,\um, the {\it Spitzer} observations of \citet{tanner2009} are also included. These seemed to confirm the ring-belt model in our earlier paper. The new PACS datum is marginally below the blackbody, but also marginally above the composite SED and, at the moment, we need to leave this issue as undecided. 

\subsection{Image de-convolution}

From high-S/N data, possible fine-structural details in the image can be retrieved. This would require image sharpening techniques, such as, e.g., image de-convolution with the known PSF.

The images were de-convolved with the rotated Arcturus-PSF using the algorithm for a maximum entropy method (MEM) developed by \citet{hollis1992}. Guided by experience, we stopped the routine after about a dozen iterations in order not to produce artefacts and/or false super-resolution. More automatic and/or quantitative methods were tried, but were eventually abandoned due to insufficient reliability. In Fig.\,\ref{observed}, the direct observations are compared with the de-convolved images and in Table\,\ref{results}, values measured on the sharpened images are reported. 

In Figure\,\ref{ring}, one-dimensional cuts along the major axis of the observed and de-convolved 100\,\um\ image are displayed. It would have been more natural to select the observation at 70\,\um, having the highest spatial resolution, but the 100\,\um\ scan map data are of considerably higher S/N, outweighing the apparent resolution advantage of the shorter wavelength data. In the analysed sub-frame, the flux was conserved within 7\% by the MEM routine. Prior to the de-convolution, a stellar point source with photospheric flux of 8\,mJy at 100\,\um\ (Table\,\ref{results}) was subtracted from the PACS image. The resulting sharpened image reveals a central {\it broadened plateau} and a central depression with a depth of about 2\%, which is consistent with the debris residing in a ring or belt around the star. 

With standard assumptions regarding the emitting grains (astronomical silicates, a blow-out size limit $a_{\rm min}=  0.6$\,\um\ for the F9\,V star and a $-3.5$ power law index for the size distribution) we find that the 100\,\um\ surface brightness profiles along the major and minor axes are well reproduced assuming a disk inclination of about 70\adeg\ and a two-parameter model for the surface density, $\Sigma(r)$. These parameters are the peak density position, $r_{\rm max}$, and the power law index of the surface density profile for $r > r_{\rm max}$. The best fit to the surface brightness profiles is consistent with a ring-like disc, having values of $r_{\rm max} \sim 85$\,AU and $\Sigma(r > r_{\rm max}) \propto r^{-3}$, respectively. A more elaborate model, with the size distribution computed self-consistently and taking into account the profiles also at other wavelengths, will be presented by Augereau et al. (in prep.).

This roughly 40\,AU wide ring or belt at about 85\,AU from the star appears similar to the EKB of the Solar System. Based on an analogy with the debris disc around Fomalhaut and on theoretical expectations, it is quite possible that another gas giant planet, \qeri\,c, could be orbiting the star inside the inner belt edge. Given the age of the system, \gapprox\,2\,Gyr, the direct detection of \qeri\,c, for instance by means of coronography, can be expected to be hard \citep[see, e.g.,][]{beichman2006}.

\section{Conclusions}

Based on imaging observations with PACS in the three photometric bands at 70\,\um, 100\,\um\ and 160\,\um\ we find that
\begin{itemize}
\item[$\bullet$] the debris around the solar-type star \qeri\ has an oval-shaped brightness distribution, the size of which increases with the wavelength; 
\item[$\bullet$] the integrated flux density at these wavelengths leads to an SED which is in good agreement with earlier results;
\item[$\bullet$] the very high signal-to-noise of the 100\,\um\ scan map is adequate to sharpen the image using an image de-convolution technique, revealing a ring-like structure with maximum surface density at \about\,85\,AU from the star;
\item[$\bullet$] with a width of about 35 to 45\,AU, this ring or belt around the F9\,V star \qeri\ is similar to the Edgeworth-Kuiper Belt around the Sun. This may hint at the presence of another planet, \qeri\,c.
\end{itemize}

\acknowledgement{We wish to acknowledge the support by the SNSB, the CNES, the PNP and the MICINN. In addition, we have benefited from HCSS / HSpot / HIPE which are joint developments by the {\it Herschel} Science Ground Segment Consortium, consisting of ESA, the NASA {\it Herschel} Science Center, and the HIFI, PACS and SPIRE consortia.}

\end{document}